**Choosing an analytic approach: Key study design considerations in state policy evaluation**

Elizabeth M. Stone[1,2], Megan S. Schuler[3], Elizabeth A. Stuart[4], Max Rubinstein[3], Max Griswold[3], Bradley D. Stein[3], Beth Ann Griffin[3]

1. Rutgers Institute for Health, Health Care Policy and Aging Research, New Brunswick, NJ, USA
2. Department of Psychiatry, Rutgers Robert Wood Johnson Medical School, New Brunswick, NJ, USA
3. RAND Corporation, Santa Monica, CA, USA
4. Department of Biostatistics, Johns Hopkins Bloomberg School of Public Health, Baltimore, MD, USA

**Corresponding author:**
Elizabeth M. Stone, PhD
112 Paterson Street, Room 305
New Brunswick, NJ 08901
elizabeth.stone@rutgers.edu

**Page count:** 31 (excluding exhibits)
**Figures/Tables:** 5

**Statements and Declarations:** The authors have no conflicts of interest to disclose.

**Acknowledgments:** This work was supported by funding from the National Institute on Drug Abuse (P50DA046351). The content is solely the responsibility of the authors and does not necessarily represent the official views of NIDA, the NIH or the US Government.

**ABSTRACT**

This paper reviews and details methods for state policy evaluation to guide selection of a research approach, based on an evaluation's setting and available data. We highlight key design considerations for an analysis, including treatment and control group selection, timing of policy adoption, expected effect heterogeneity, and data considerations. We then provide an overview of analytic approaches and differentiate between methods based on an evaluation's context, such as settings with no control units, a single treated unit, multiple treated units, or with multiple treatment cohorts. Methods discussed include: interrupted time series models, difference-in-differences estimators, autoregressive models, and synthetic control methods, along with method extensions which address issues like staggered policy adoption and heterogenous treatment effects. We end with an illustrative example, applying the developed framework to evaluate the impacts of state-level naloxone standing order policies on overdose rates. Overall, we provide researchers with an approach for deciding on methods for state policy evaluations, which can be used to select study designs and inform methodological choices.

## 1 INTRODUCTION

Under American federalism, states (rather than the federal government) hold much of the power for creating laws and policies that relate to public health, public safety, and social well-being (Greer et al., 2023; Rich & White, 1997). Over the past 15 years, partisanship and political polarization have led to states enacting more heterogeneous, divergent public health policies including those related to Medicaid financing, the COVID-19 pandemic, and the ongoing opioid crisis (Birkland et al., 2021; Oberlander, 2020). The "natural experiments" resulting from state-level variation in policy adoption can be leveraged by health and social policy researchers to assess the consequences, both intended and unintended, of various state policies across a wide range of outcomes (Aiken et al., 2022; Schell et al., 2020; Schuler et al., 2020; Zhang & Warner, 2020). As these studies may impact future policy decisions, there is a critical need for high-quality, rigorous policy evaluations.

Applied researchers face many potential challenges when designing evaluations including issues related to the study design and analytic approach (Rudolph et al., 2022; Schuler et al., 2021; Smart et al., 2020). For example, state policy evaluations can often face challenges due to sample size and power since there are only 50 states, only a small number of states might be treated with a policy, or polices that appear similar on their face may have differences in included provisions, implementation, or enforcement that don't allow them to be meaningfully grouped together (Grant et al., 2020; Schuler et al., 2020). Even when states do adopt similar policies, they may enact them at different times (i.e., "staggered adoption") or in conjunction with other policies. Both scenarios introduce bias if an approach does not account for time trends or concurrent policies (Griffin et al., 2023a; Matthay et al., 2021).

Another challenge for policy evaluation is that policies are not randomly assigned and may select into treatment. States that choose to adopt a policy are likely to differ from states without the policy on important characteristics, including the intended target of the policy; for example, states with high overdose rates may be more likely to pass policies aimed at reducing overdose deaths (Griffin et al., 2023b; Lurie & Sharfstein, 2023). Policymaking regarding many public health topics continues to exhibit strong partisan trends across states – e.g., states that expanded Medicaid under the ACA generally look systematically different in terms of generosity of public health and social welfare policies than states that did not expand Medicaid (Oberlander, 2020). Additionally, there is growing attention to potential heterogeneity in policy effects across states, across subpopulations within a state, and across time (Dave et al., 2021; Wing et al., 2018).

In recent years, there has been a proliferation of methods aiming to address these challenges in policy research (Baker et al., 2022; Degli Esposti et al., 2020; Roth & Sant'Anna, 2023; Wing et al., 2018). In this quickly evolving landscape, understanding and selecting approaches to analyze a research question can be daunting. In this paper, we aim to aid researchers by 1) outlining key study design features to consider when selecting the methods for a state policy evaluation, 2) providing an overview of commonly used methods given these design features, and 3) presenting an illustrative example considering these key study design features in the context of a state opioid policy evaluation.

## 2 EVALUATION CONTEXT and ASSUMED DATA STRUCTURE

In our discussion, we focus on methods for estimating the impact of a single policy implemented at the state level in at least one state, using annual longitudinal data of policy exposure, outcomes, and covariates at the state-level. These considerations also apply more broadly when using measures on other time scales (e.g., monthly) or geographies (e.g., counties).

To formalize the setting and inferential goal, we use potential outcomes notation for repeated measures data. Let $Y_{it1}$ denote the potential outcome (e.g., opioid-related mortality rate) for state $i$ ($i = 1, \dots, 50$) if the policy was in effect at time *t* and let $Y_{it0}$ denote the potential outcome for state $i$ if the policy was not in effect. Each state thus has two potential outcomes at each time point – only one of which is observed – while the other represents the counterfactual (e.g., for treated states, what would have occurred in the absence of the policy; for control states, what would have occurred had the policy been implemented).

For the methods presented in this paper, the estimand of interest is the average treatment effect on the treated (ATT), either overall or decomposed for subgroups (e.g., group-time). Specifically, the ATT is defined as $ATT = E[\, Y_{it1} - Y_{it0} \mid D_i = 1 \,]$, where $D_i = 1$ indicates states that adopted the policy. Because $Y_{it1}$ is observed for treated states, ATT inference requires estimating the unobserved counterfactual $Y_{it0}$. The methods we discuss differ in how they estimate this counterfactual $Y_{it0}$, relying on distinct assumptions and model specifications. To guide applied researchers, we present the underlying causal assumptions and key sensitivity analyses for each method. Although the ATT is the primary estimand across most approaches, variation in model specification can yield different definitions or decompositions of the ATT (e.g., subgroup- or time-specific ATTs). We note that some alternative methods in the broader literature target different estimands, such as the average treatment effect (ATE) or the average treatment effect on the untreated (ATU), which require different strategies for approximating counterfactuals; these methods are beyond the scope of this discussion.

## 3 KEY STUDY DESIGN FEATURES and ANALYTIC DECISIONS

In this section, we outline key considerations to guide researchers in selecting methods for a given study. A critical first step is assessing the data available for analyses (and its limitations), ensuring that outcome measures accurately capture the constructs of interest and that the policy of interest is appropriately operationalized (Barsky et al., 2025; Smart et al., 2020). Researchers should also consider the impact of model misspecification on inferences, as misspecification can meaningfully alter estimated policy effects.

### 3.1 Definition of the Treated Group

A first consideration is determining which states are categorized as treated. This can vary based on the number of states and timing of policy adoption:

- *Single treated state:* In some cases, there may be only one state that adopts a given policy (or does so within the study period).
- *Multiple treated states:* When several states adopt a policy, researchers should carefully assess whether states with different policy provisions or implementation approaches can be reasonably grouped together and considered the "same" policy (see Grant et al. (2020); Schuler et al. (2021); Schuler et al. (2023) for further discussion). Timing of adoption is also critical for informing the analytic approach:
    - *Simultaneous policy adoption:* If there is a single treatment cohort (i.e., more than one state adopts a policy and all treated states adopt the policy at the same time), this is considered simultaneous policy adoption or "common timing" (Wooldridge, 2021).
    - *Staggered policy adoption:* More commonly, states adopt the policy at different times, creating multiple "treatment cohorts" defined by year of adoption (Callaway & Sant'Anna, 2021). Under staggered adoption, states differ in their duration of policy exposure, based on the date of policy adoption (Wooldridge, 2021).

### 3.2 Definition of the Comparison Group

As with the treated group, researchers must define the control group, which depends on available data and policy adoption patterns:

- *No comparison group:* In some settings, a control group may not be feasible – for example, when data are available for only a single state or when a federal policy is enacted simultaneously across all states.
- *Comparison group:* When creation of a control group is feasible, selection of control states should be informed by data availability, analytic method, and study-specific inclusion and exclusion criteria. In some cases, control states may be restricted to those with characteristics that improve comparability to treated states – for example, states with a similar policy environment prior to adoption of the policy of interest (see Seewald et al. (2024)). In staggered adoption designs, a central methodological consideration is whether the comparison group includes only "never treated" states (i.e., those that never adopt the policy) or also "not-yet-treated" states during their pre-policy period. While the latter approach can improve efficiency, it may introduce bias by conditioning on future treatment status (Callaway & Sant'Anna, 2021; Hernán & Robins, 2016; Seewald et al., 2024).

### 3.3 Key Model Assumptions

The specific assumptions required by each design and analytic approach vary; here we summarize some of the more common and fundamental assumptions:

- *Ignorability:* Treatment assignment and observed outcomes are not a function of unobserved, time-varying confounders (see further discussion of confounders below).
- *Positivity:* There is overlap in the distribution of covariates between the treated and control groups.
- *No anticipation:* The policy does not affect outcomes prior to its adoption (i.e., there is no pre-emptive behavior change in anticipation of the policy).
- *Consistency:* The potential outcome under the observed treatment assignment is equal to the observed outcome. This assumption requires clear definition of the treatment assignment (i.e., presence of a policy) and is more likely to hold with homogenous policy groups.
- *No spillover effects:* One state's treatment status does not impact outcomes in another state. This may be of particular concern for neighboring states.
- *Parallel (counterfactual) trends:* Specific to difference-in-differences models, this assumption requires that had the treated state(s) not adopted the policy, the outcome in the treatment state(s) in the post-treatment period would follow the same trend observed in the control state(s) into the post-period. Because post-period outcomes under the control condition are unobserved, this assumption cannot be tested directly. Pre-period "parallel trends" tests may provide suggestive evidence but cannot definitively establish validity.

### 3.4 Anticipated Policy Effect Heterogeneity

Policy effects may vary across states, over time, or by timing of policy adoption. Investigating this heterogeneity is an important early step in study design. Before estimating formal evaluation models, researchers should conduct exploratory descriptive analyses, such as examining outcome trajectories stratified by state, treatment cohorts, or timing of policy adoption, to gauge whether effect heterogeneity is likely. These insights can guide the choice of analytic approach, including whether to use models that estimate time-varying or cohort-specific effects.

- *Effect heterogeneity over time:* Even for a single treated state, policy effects may evolve over time after implementation. In such cases, methods that estimate time-varying effects, rather than a single average post-policy estimate, are preferable. Depending on the research question, these dynamic effects may serve as the primary estimand or as contextual findings along with primary effect estimates.
- *Effect heterogeneity by treatment cohort:* When multiple states adopt a policy at different times, effects may differ across cohorts. For example, later adopters may achieve larger policy impacts after learning from early adopters' implementation experiences. Researchers may select methods that estimate cohort-specific effects in addition to an overall policy effect, or may conduct separate analyses for each cohort. As with dynamic effects, reporting cohort-specific effects may be central to the study aims or may be included as contextual findings.

### 3.5 Data Considerations

The quality and structure of available data are central to selecting an appropriate analytic approach. Below we highlight several practical considerations relevant to policy evaluation (for additional discussion in the opioid policy evaluation context, see Schuler et al., 2021 and Pacula et al., 2025):

- *Number of repeated measures:* Policy evaluations generally rely on repeated outcome measures. The amount of pre- and post-policy data to include depends on study design, availability, and methodological approach. In settings with staggered adoption, states contribute varying numbers of pre- and post-policy observations. Researchers must decide whether to use a "balanced panel," limited to the pre- and post-policy periods observed for all states, or an "unbalanced panel" that uses all available data, acknowledging that some time points will include data from only a subset of states.
- *Data completeness:* Some methods (e.g., synthetic control approaches) cannot accommodate missingness in the repeated measures data. Strategies for addressing missingness may depend on timing (i.e., pre- vs post-policy) and treatment status (i.e., treated vs. control) of the missing observations. For example, researchers may exclude control states with missing outcome data or impute missing outcome data for a treatment state or pre-policy observation.
- *Policy treatment cohorts:* When multiple states adopt a policy, treatment cohorts are often defined by adoption timing (e.g., all states adopting the policy in the same year when using annual data). Ideally, there would be multiple states per cohort and numerous cohorts, though in practice researchers must balance these ideals with available data.
- *Definition of policy and outcome measures:* Policy evaluations commonly use a binary policy indicator and continuous outcome, but methods differ in flexibility to accommodate various types of measures (e.g., categorical policy exposure). Additionally, in defining both policy and outcome measures, researchers should assess sensitivity to policy specification – such as how adoption timing, provisions, or implementation stages are defined – as these choices can substantially affect estimated policy effects. Sensitivity analyses, such as assessing how robust findings are to redefining the policy variable based on specific law provisions or implementation features, can help assess whether grouping policies into a single measure introduces bias.

### 3.6 Additional Considerations

Two additional considerations, confounding and relative model performance, may also guide selection among viable analytic approaches for a given evaluation.

- *Confounding:* Identifying potential confounders in a policy evaluation context depends on the underlying causal assumptions specific to the chosen approach (Zeldow & Hatfield, 2021). Once identified, confounders can be addressed through study design decisions (e.g., restricting control units to those with similar characteristics) or analytic strategies (e.g., including covariates in outcome models or weighting control units based on observed baseline characteristics).
- *Relative model performance:* In each evaluation context, analytic approaches vary in performance measures including bias, type I error rates, and coverage. Simulation studies and tools to compare approaches on these metrics may help researchers make informed decisions on choice of analytic approach when multiple methods may be appropriate (Griffin et al., 2025; Griffin et al., 2023a; Griffin et al., 2023b; Griffin et al., 2021).

## 4 METHODS FOR POLICY EVALUATION

In the following sections, we detail analytic methods for state policy evaluations, with respect to the key design features highlighted above. Specifically, we differentiate between methods for the following settings: (1) *single treated state/cohort, no comparison state(s)*, (2) *multiple treatment cohorts (staggered adoption), no comparison state(s)*, (3) *single treated state, multiple comparison states*, (4) *single treated state/cohort (simultaneous adoption), with comparison state(s)*, and (5) *multiple treated states/cohorts (staggered adoption), with comparison state(s)* (Figure 1). Table 1 provides brief summaries of each method.

### *4.1 Methods for a single treated state (or single treatment cohort), no comparison state(s)*

#### *4.1.1 Interrupted Time Series (ITS)*

In the interrupted time series (ITS) model, repeated measures over time (the “time series”) are used to assess changes in magnitude and trend of outcomes before and after a policy is enacted (the “interruption”). In its most basic form, the ITS model does not require a control group. The ITS model is expressed as:

$$g\left(Y_{it}^{obs}\right) = \beta_0 + \beta_1 time_t + \beta_2 policy_t + \beta_3 time_since_policy_t + \rho_i + \epsilon_{it} \quad (1)$$

where $g(.)$ denotes the generalized linear model (GLM) link function (e.g., linear), $\rho_i$ denotes state fixed effects (in the case of multiple states in a single treatment cohort), and $\varepsilon_{it}$ denotes the error term. The measures of interest in this model are $time$, which measures time elapsed since the start of the study period, $policy$, a time-varying indicator that denotes the policy is in effect at time $t$, and $time_since_policy$, which measures time elapsed since the policy was implemented. In this model, the coefficients of interest are $\beta_2$, which indicates the immediate change in outcome at the time of interruption (change in level), and $\beta_3$, which indicates the change in outcomes over time following vs. before the interruption (change in slope).
**Underlying causal assumption:** The outcome in the treated state would have continued uninterrupted in both level and trend if not for the policy.
**Key sensitivity analyses:** placebo test of treatment timing (i.e., test pseudo-treatment time in the observed pre-period); placebo test of outcome (i.e., test policy effect on outcome not expected to be impacted); sensitivity to model specifications
**Model assumption(s):** Ignorability, no anticipation, consistency
**Effect heterogeneity by time:** Effect heterogeneity over time is captured in $\beta_3$ which indicates the change in outcomes over time following policy adoption.
**Effect heterogeneity by treatment cohort:** Not applicable – single treatment state/cohort (effects assumed to be homogenous within treatment cohort)
**Data consideration(s):** Number of repeated measures, policy and outcome definition
**References:** Berk (2022); Bernal et al. (2017); Ewusie et al. (2020); Kontopantelis et al. (2015)

### *4.2 Methods for multiple treatment cohorts (staggered adoption), no comparison state(s)*

#### 4.2.1 ITS with multiple baselines

One extension of the basic ITS design is ITS with multiple baselines. In cases of staggered treatment adoption, there are different baseline periods for the different treatment cohorts. This staggered adoption allows for the estimation of effects in different states, with different baseline trends, and at different points in calendar time which help to account for temporal trends and other changes taking place at the time of the intervention. In ITS with multiple baselines, separate ITS models are fit for each treatment cohort and effects from all models can then be averaged.
**Underlying causal assumption:** The outcome in the treated states would have continued uninterrupted in both level and trend if not for the policy.
**Key sensitivity analyses:** placebo test of treatment timing; placebo test of outcome; approach to effect aggregation (e.g., simple average vs. weighted average); sensitivity to model specifications
**Model assumption(s):** Ignorability, no anticipation, consistency, no spillover effects
**Key data considerations:** Sufficient pre-period data to model outcomes over time
**Effect heterogeneity by time:** Effect heterogeneity over time is captured by the coefficient indicating the change in outcomes over time following policy adoption.
**Effect heterogeneity by treatment cohort:** Effect heterogeneity by treatment cohort can be assessed by comparing the separate results for each cohort; heterogeneity by treatment cohort is not formally tested.
**Data consideration(s):** Number of repeated measures, policy and outcome definition
**References:** Biglan et al. (2000); Hawkins et al. (2007)

### *4.3 Methods for a single treated state, with more than one comparison state*

*4.3.1 Classic synthetic control method (SCM)*

The synthetic control approach is used to assess the impact of an intervention or policy change on a single state. This approach involves constructing a single "synthetic" control group (a weighted combination of control states) that matches the treated state as closely as possible on the outcome trends and potential confounders during the pre-policy period (Abadie et al., 2010). When creating the synthetic control group, states that are most similar to the policy state in the pre-policy period are "upweighted" (receive the largest weights) and states that are more dissimilar are "downweighted." The effect estimate is then calculated as the difference between the treated group and the synthetic control in the post-policy period. Because SCM does not use an outcome model, p-values are calculated based on differences between observed values and permutation tests estimating placebo effects with control states. In contrast to the difference-in-differences designs (discussed below), SCM can explicitly account for time-varying confounders, as the weights match the treated and synthetic control group across the pre-policy period (Kreif et al., 2016).

**Underlying causal assumption:** The outcome trend and level in the synthetic control group in the post-period is what would have been observed in the treated state if not for the policy.
**Key sensitivity analyses:** placebo test of treatment timing; selection of control group
**Model assumption(s):** Ignorability, positivity, no anticipation, consistency, no spillover effects
**Effect heterogeneity by time:** Time-specific effects are automatically estimated in SCM programs.
**Effect heterogeneity by treatment cohort:** Not applicable – single treated state
**Data consideration(s):** Number of repeated measures, policy and outcome definition
**References:** Abadie (2021); Abadie et al. (2010); Abadie and Gardeazabal (2003); Abadie and L'Hour (2021)

*4.3.2 Augmented synthetic control method (ASCM)*

The augmented synthetic control method is an extension of the traditional SCM designed for settings in which traditional SCM's weighting does not achieve satisfactory matching of the treated and control units in the pre-policy period (Ben-Michael et al., 2021). ASCM modifies the traditional SCM approach by 1) adding an outcome model to adjust for any remaining pre-treatment imbalances in outcomes or covariates between the treated state and the synthetic control and 2) allowing for negative weighting of control states, to provide better similarity in the pre-period. If traditional SCM achieves satisfactory weighting, results from traditional SCM and ASCM will be similar.

**Underlying causal assumption:** The outcome trend and level in the synthetic control group in the post-period is what would have been observed in the treated state if not for the policy.
**Key sensitivity analyses:** placebo test of treatment timing; selection of control group; sensitivity to model specifications
**Model assumption(s):** Ignorability, positivity, no anticipation, consistency, no spillover effects
**Effect heterogeneity by time:** Time-specific effects are automatically estimated in ASCM programs.
**Effect heterogeneity by treatment cohort:** Not applicable – single treated state
**Data consideration(s):** Number of repeated measures, policy and outcome definition
**References:** Ben-Michael et al. (2021)

***4.4 Methods for single treated state or treatment cohort (simultaneous adoption), with comparison state(s)***

*4.4.1 "Classic" (two-way fixed effect) difference-in-differences model (DID)*

A common model in state policy evaluations is the classic two-way fixed effect difference-in-differences model (Dimick & Ryan, 2014; Wing et al., 2018). DID has a long history in policy evaluation, harkening back to John Snow's study of cholera in 1855 (Snow, 1855). The DID estimate essentially subtracts the observed pre-policy to post-policy change in the comparison group from the observed pre-policy to post-policy change in the policy group, hence the name "difference-in-differences." The classic DID specification is often implemented as a two-way fixed effects model that includes both state- and time-fixed effects expressed as:

$$g(Y_{it}^{obs}) = \beta_0 + \beta_1 trt_i + \beta_2 policy_t + \beta_3 (trt) \times (policy)_{it} + \rho_i + \sigma_t + \varepsilon_{it} \quad (2)$$

where $g(.)$ denotes the generalized linear model (GLM) link function (e.g., linear) and $\varepsilon_{it}$ denotes the error term. $trt$ is a (time-invariant) indicator whether a given state is ever treated, $policy$ is a time-varying indicator that denotes the policy is in effect at time $t$, and $(trt) \times (policy)$ term is the interaction of the two. $\beta_3$, the coefficient of the interaction term, is the estimate of the treatment effect. State fixed effects, $\rho_i$, quantify baseline differences in the outcome across states and time fixed effects, $\sigma_t$, quantify national temporal trends. State fixed effects only account for time-invariant differences between states and time fixed effects only account for exogenous factors that affect both treated and untreated states equally.
**Underlying causal assumption:** If not for the policy, the treated state(s) would exhibit the same average change in the outcome from pre- to post-policy as was observed in the control state(s).
**Key sensitivity analyses:** selection of control group; sensitivity to model specifications
**Model assumption(s):** Positivity, no anticipation, consistency, no spillover effects, parallel trends
**Effect heterogeneity by time:** The event study/dynamic DID approach described below can be used to estimate time-specific treatment effects.
**Effect heterogeneity by treatment cohort:** Not applicable – single treated state or treatment cohort (effects assumed to be homogenous within treatment cohort)
**Data consideration(s):** Policy and outcome definition
**References:** Abadie (2005); Baker et al. (2025); Bertrand et al. (2004); Chabé-Ferret (2017); Daw and Hatfield (2018a, 2018b); Ryan (2018); Ryan et al. (2015); Stuart et al. (2014); Zeldow and Hatfield (2021)

*4.4.2 Event Study/Dynamic DID*

The classic DID model generates a single point estimate for the policy effect, representing the average effect across the observed post-policy period. A notable and important extension to the classic DID model is the event study design, which has been employed in the economics literature since the 1930s and allows for estimation of the time-varying effect of a policy (de Chaisemartin & D'Haultfœuille, 2020). Essentially, an event time study defines time 0 as the time of policy implementation and examines time-specific treatment effects relative to a given time point (typically the time period immediately preceding policy adoption). The post-policy period is indexed by positive numbers ($k = 1, \dots, T_1$, where $T_1$ represents the maximum number of time periods observed in the post-period) and accounted for in the model by "lagging indicators." Inclusion of lagging indicators allows estimation of time-specific effect estimates in the post-policy period, thereby relaxing the classic DID assumption that the treatment effect is constant over time.

An event study model could also include "leading indicators" which span the pre-policy period (generally indexed by negative numbers $k = -T_0, \dots, -1$, where $T_0$ is the maximum number of time periods observed in the pre-period). Inclusion of these leading indicators requires extending the common trends assumption to also hold for the pre-policy period in addition to the post-policy period (Wing et al., 2018).

A full "DID event study" (or "dynamic DID") includes the complete set of both leading and lagging indicators. The general form for a DID event time study is as follows:

$$g(Y_{it}^{obs}) = \beta_0 + \sum_{k=T_0}^{-2} \beta_k (1(t=k) \cdot policy_{ik}) + \sum_{k=0}^{T_1} \beta_k (1(t=k) \cdot policy_{ik}) + \beta X_{it} + \rho_{\boldsymbol{i}} + \sigma_t + \varepsilon_{it} \quad (3)$$

where $1(t = k)$ is an indicator that equals 1 if the observation's event time indexed time is equal to $k$ and 0 otherwise. The lagging indicators comprise the summation term indexed $k = 1, \dots, T_1$ and leading indicators comprise the summation term indexed $(T_0, \ldots, -2)$. To avoid multicollinearity, one period is dropped (traditionally $T = -1$).
**Underlying causal assumption:** If not for the policy, the treated state(s) would exhibit the same change in the outcome trends from pre- to post-policy as was observed in the control state(s).
**Key sensitivity analyses:** selection of control group; sensitivity to model specifications
**Model assumption(s):** Ignorability, positivity no anticipation, consistency, no spillover effects, parallel trends
**Effect heterogeneity by time:** This model estimates time-specific treatment effects for each time point.
**Effect heterogeneity by treatment cohort:** Not applicable – single treated state or treatment cohort (effects assumed to be homogenous within treatment cohort)
**Data consideration(s):** Number of repeated measures, policy and outcome definition
**References:** Freyaldenhoven et al. (2021); Miller (2023)

*4.4.3 Comparative Interrupted Time Series (CITS)*

The comparative interrupted time series model is another extension of the basic ITS design which adds a comparison group. Conceptually the CITS design is similar to DID, though CITS requires more years of data. The basic CITS model extends the ITS model to include measures indicating treatment vs. control states and interactions of the treatment indicator with the time, policy, and time_since_policy variables:

$$g\left(Y_{it}^{obs}\right) = \beta_0 + \beta_1 time_t + \beta_2 policy_t + \beta_3 time_since_policy_t + \beta_4 treatment_i + \beta_5 trtXtime_{it} + \beta_6 trtXpolicy_{it} + \beta_7 trtXtime_since_policy_t + \beta X_{it} + \rho_i + \epsilon_{it} \quad (4)$$

In this model, the coefficients of interest are $\beta_6$, which indicates the difference in the immediate change in outcome at the time of interruption (change in level) between treatment and control states, and $\beta_7$, which indicates the difference in change in outcomes over time following vs. before the interruption (change in slope) between treatment and control states. By comparing the trend of the outcome between the states that receive the policy change those that do not, the CITS model can estimate the magnitude and direction of the policy effect.
**Underlying causal assumption:** If not for the policy, the post-policy outcomes in the treated state(s) would have evolved in the same way as was observed in the control state(s).
**Key sensitivity analyses:** selection of control group; sensitivity to model specifications
**Model assumption(s):** Positivity, no anticipation, consistency, no spillover effects
**Effect heterogeneity by time:** Time-specific effects can be estimated using the CITS model.
**Effect heterogeneity by treatment cohort:** NA – the basic CITS model assumes simultaneous policy adoption (i.e., a single interruption). CITS models can be extended to account for multiple interruptions, though this is typically for multiple policy changes in a treated state or treatment group rather than for staggered policy adoption.
**Data consideration(s):** Number of repeated measures, policy and outcome definition
**References:** Fry and Hatfield (2021); Lopez Bernal et al. (2018)

***4.5 Methods for multiple treated states or treatment cohorts (staggered adoption), with comparison state(s)***

*4.5.1 DID extensions: Cohort-based DID extensions and Imputation-based DID extensions*

Until fairly recently, the issue of staggered adoption – present in most state policy evaluations – was not given particular attention, as the classic DID model can (mathematically) handle this situation and provide a policy estimate. However, recent methodological work has highlighted that effect estimates from classic DID models may be biased in the presence of staggered adoption if policy effect heterogeneity exists (Borusyak et al., 2024; Callaway & Sant'Anna, 2021; de Chaisemartin & D'Haultfœuille, 2020; Goodman-Bacon, 2021; Imai & Kim, 2021; Sun & Abraham, 2021). In the presence of staggered adoption, there are distinct “pre-policy” and “post-policy” periods with respect to each treated state that should be addressed. It becomes less clear which states should comprise the control group for a given treated state (e.g., only never-treated states? Or also not-yet-treated states?) and the models can inadvertently adjust for what are essentially “post-treatment” outcomes, which can lead to bias.

In particular, Goodman-Bacon (2021) showed that, in the context of staggered adoption, the classic DID is comprised of the weighted average of *all possible comparisons* of treated and control states. For example, assume that there are two groups of treated states, “early adopters” and “late adopters” in addition to a comparison group that never implements the policy of interest. There are now multiple possible comparisons: early adopters vs. untreated and late adopters vs. untreated, as well as the early adopters vs. late adopters (before the late adopter group implemented policy) and late adopters vs. early adopters (after the early adopter group implemented policy). Methodological work has characterized the latter two contrasts as “forbidden” contrasts that should be excluded as they include comparisons between groups that have already been treated, but initiated treatment at different times (Goodman-Bacon, 2021). Furthermore, it has been shown that the classic DID estimator will only yield an unbiased estimate in the context of staggered adoption if the treatment effect is homogenous across states and across time (de Chaisemartin & D'Haultfœuille, 2020). Systematic differences between states that adopt (vs. do not adopt) a policy and changes in policy implementation or enforcement over time that can impact the policy impacts (e.g., ramp up period to full implementation) make the assumptions of homogeneous treatment effects highly unlikely in most policy evaluations (Schuler et al., 2021). Multiple DID-based methods, such as those detailed below, have been recently developed specifically to handle staggered adoption and heterogeneous treatment effects.

One genre of DID extensions essentially creates a series of cohorts of states who implemented the policy at the same time, conducts a DID for each (having eliminated the issue of staggered adoption by re-anchoring time for each cohort), and then aggregates the cohort-specific estimates in various ways to summarize the overall policy effect. Having calculated these estimates for each group and time period, the group-time treatment effects are then aggregated to form an overall estimate of the treatment effect. Examples of cohort-based DID extensions include methods by Callaway and Sant’Anna (2021); de Chaisemartin and D'Haultfœuille (2020); Roth and Sant’Anna (2023); and Sun and Abraham (2021).

Another category of DID-based extensions can be termed imputation-based methods. For example, a method proposed by Borusyak et al. (2024) uses all untreated observations (i.e., all observations from “never treated” states + pre-policy observations from “not yet treated” states) to estimate a classic two-way fixed effect DID model. Using this untreated sub-sample estimates a “counterfactual” for each treated unit in the absence of treatment. Next, the treatment effect is calculated as the (population-weighted) average of the difference between the observed outcome and the predicted counterfactual, $Y_{it}^{obs} - \hat{Y}_{it}$ . These approaches yield valid estimates when the parallel trend assumption holds for all groups and time periods and there is no anticipation effect.

**Underlying causal assumption:** If not for the policy, the treated state(s) would exhibit the same average change in the outcome from pre- to post-policy as was observed in the control state(s).
**Key sensitivity analyses:** selection of control group; sensitivity to model specifications
**Model assumption(s):** Positivity, no anticipation, consistency, no spillover effects, parallel trends
**Effect heterogeneity by time:** These cohort- and imputation-based approaches can easily be conducted in a way that estimates time-specific treatment effects.
**Effect heterogeneity by treatment cohort:** Somewhat by definition, the cohort-based analyses estimate cohort-specific treatment effects. Imputation-based DID extensions do not use treatment cohorts but individual DID models can be estimated for each treated state or treatment cohort.
**Data consideration(s):** Number of repeated measures, policy and outcome definition, policy treatment cohorts
**References: Overview of DID extensions:** Roth et al. (2023); Wang et al. (2024); **Cohort-based DID extensions:** Callaway and Sant’Anna (2021); de Chaisemartin and D'Haultfœuille (2020); Roth and Sant'Anna (2023); Sun and Abraham (2021); **Imputation-based DID extensions:** Borusyak et al. (2024); Gardner (2022); Liu et al. (2024); Powell et al. (Under Review)

*4.5.2 Debiased autoregressive (AR) models*

Debiased autoregressive models are another class of methods recently highlighted as promising for policy evaluation (Antonelli et al., 2024). Standard autoregressive models include one or more lagged measures of the outcome variable (e.g., $Y_{it-1}^{obs}$) as covariates. However, when estimating causal effects, incorporating lagged outcomes into models can lead to biased effect estimation when the lagged outcomes capture parts of the policy effect (Griffin et al., 2021; Schell et al., 2018). A recently proposed solution to this problem are so-called "debiased autoregressive models." These remove effects of the policies from prior periods as a strategy to obtain unbiased causal effects while controlling for potential confounding from differences in prior outcome trends absent treatment across treated and comparison states. One parameterization of a debiased AR model with a single lagged value of the outcome expressed as $Y_{it-1}^{obs}$ is:

$$g\left(Y_{it}^{obs}\right) = \beta_0 + \beta_1(Y_{i,t-1} - \gamma\ policy_{i,t-1}) + \gamma\ policy_{it} + \beta X_{it} + \sigma_t + \epsilon_{it} \quad (5)$$

Like the classic DID model, this model includes time fixed effects that capture temporal trends. However, the model also adjusts for state-specific variability using the debiased AR term ($\beta_1 \cdot (Y_{it-1} - \gamma\ policy_{i,t-1})$) rather than state fixed effects. Notice that in a setting where states are only treated in the final time period, $policy_{i,t-1}$ is zero for all units, so that in this setting this becomes a standard AR model.

**Underlying causal assumption:** The policy is effectively randomized at every time point conditional on the covariates and the prior outcomes absent the policy.

**Key sensitivity analyses:** specification of lagged term(s); selection of control group; sensitivity to model specifications

**Model assumption(s):** Ignorability (conditional on covariates and prior outcomes absent treatment), positivity, no anticipation, consistency, no spillover effects

**Effect heterogeneity by time:** This model can utilize an event study/dynamic approach similar to DID to estimate time-specific treatment effects.

**Effect heterogeneity by treatment cohort:** This model can be used to estimate cohort-specific treatment effects by directly including both the main effects of cohorts as well as cohort interaction effects in the model.

**Data consideration(s):** Number of repeated measures, policy and outcome definition

**References:** Antonelli et al. (2024); Schell et al. (2018)

*4.5.3 ASCM extension for staggered adoption:*

The augmented synthetic control method can be extended for settings with staggered treatment adoption. In this approach, the synthetic control units are weighted to minimize a weighted average of pooled and unit-specific pre-treatment fits. This combination can range from 0, with separate synthetic control weights estimated for each treatment unit, to 1, a fully pooled synthetic control weighted to estimate the mean across all treated units; a partially pooled approach is recommended (Ben-Michael et al., 2022).

**Underlying causal assumption:** The outcome trend and level in the synthetic control group in the post-period is what would have been observed in the treatment cohorts if not for the policy.

**Key sensitivity analyses:** placebo test of treatment timing; selection of control group; sensitivity to model specifications

**Model assumption(s):** Ignorability, no anticipation, consistency, no spillover effects

**Effect heterogeneity by time:** Time-specific effects are automatically estimated in ASCM programs.

**Effect heterogeneity by treatment cohort:** The program used to implement this method can also estimate effects by treatment cohorts (a separate synthetic control is generated for each cohort).

**Data consideration(s):** Number of repeated measures, policy and outcome definition, policy treatment cohorts

**References:** Ben-Michael et al. (2022)

## 5 CASE STUDY APPLICATION

We now illustrate the application of the identified decision points to an illustrative case study within the context of the opioid crisis. Access to naloxone as an overdose reversal drug is an essential tool in reducing opioid overdose deaths. One strategy states have taken to increase access to naloxone is the implementation of naloxone standing orders, which allow anyone to be dispensed and to carry naloxone regardless of whether that person has an individual naloxone prescription from a provider. In this case study, we use annual, state-level data from 1999 to 2017 for all states and Washington DC to estimate the effects of state naloxone standing order policies on total overdose deaths. As shown in Table 2, numerous states adopted standing order naloxone laws beginning in 2010. By 2017, 44 of the 50 states plus Washington DC had a standing order policy in place.

In this case study, we use national longitudinal data with measures on multiple treated and control states (Table 3). Policy adoption is staggered with states implementing standing orders between 2010 and 2017. As the underlying overdose crisis, including the proliferation of fentanyl in the drug supply and the adoption of federal and state policies related to the crisis changed significantly over this time period, it is likely that there is treatment effect heterogeneity by both treatment cohort and time (Ciccarone, 2021; Jones et al., 2019). Given multiple treatment and comparison states and the presence of staggered policy adoption, using this framework, we identify four viable methods for this analysis: cohort-based DID, imputation-based DID, AR, and ASCM extension for multiple units. Policy effect estimates from each of the four approaches are included in Table 4. While the effect estimates vary in direction (range: -1.32, 2.32), all four estimates are small in magnitude and not statistically significant. Differences in estimates are due, in part, to the different approaches taken in each method to approximate a counterfactual in estimation of the ATT. The range of these estimates may also be due to exclusion of treatment cohorts with only one state in the cohort-based DID analysis and differences in performance on variance and bias between approaches (Griffin et al., 2023a; Griffin et al., 2023b; Griffin et al., 2021). Data and code for implementing these methods as well as the other methods discussed in this paper are included in the Appendix.

## 6 DISCUSSION

In this paper we provide a resource for applied researchers to understand the most appropriate set of methods to use for a given study. We outline key study considerations based on definitions of treated and comparison groups, policy timing and heterogeneity of effects, key model assumptions, and data considerations. In most state-level evaluations with multiple treatment and control states and staggered policy adoption, researchers will still have multiple viable options to choose between. At that point, simulation studies, through which researchers can compare bias, variance, and coverage of different methods, can help determine which method might be optimal (Griffin et al., 2025; Griffin et al., 2023a; Griffin et al., 2023b; Griffin et al., 2021). As power is highly constrained in this setting -- we often have 50 or fewer states for analyses -- it is useful to understand which method or methods might have the greatest precision prior to running an outcome model. Simulation studies have been used in both the opioid and gun policy contexts to compare several leading methods including DID, ASCM, cohort-based DID, and the AR model with varying results depending on the outcome stream (Griffin et al., 2023b; Griffin et al., 2021; Schell et al., 2018). The 'optic' R library is one accessible resource for researchers interested in running such simulations (Griffin et al., 2025).

This study focused on design features but there are other key aspects of policy evaluations that require careful consideration – including the quality and limitations of the data and the impact of those on inferences from a specific study. To improve policy evaluations, researchers need to ensure outcomes and policies are measured and modeled accurately (Pacula et al., 2025). Creating more rigorously designed studies will help ensure policy makers are given access to better information when making decisions about whether to enact a particular policy in their state.

While this paper aimed to provide an overview of commonly used policy evaluation methods, evaluation methods are rapidly evolving. Future methods advancements can be seamlessly added to this decision framework as they become available. Future research is needed to fully understand the potential of each of these methods to address other scenarios that are common in policy, but present challenges in evaluations such as co-occurring policies.

## 7 CONCLUSION

Policy evaluations are often complex and many different methodological approaches exist. We outline key considerations related to data structure, policy adoption, and modeling assumptions to help applied researchers identify optimal methods for their work.

## REFERENCES

Abadie, A.: Semiparametric difference-in-differences estimators. Rev. Econ. Stud. 72(1), 1-19 (2005) doi:10.1111/0034-6527.00321

Abadie, A.: Using synthetic controls: Feasibility, data requirements, and methodological aspects. J. Econ. Lit. 59(2), 391-425 (2021) doi:10.1257/jel.20191450

Abadie, A., Diamond, A., & Hainmueller, J.: Synthetic control methods for comparative case studies: Estimating the effect of California's tobacco control program. J. Am. Stat. Assoc. 105(490), 493-505 (2010) doi:10.1198/jasa.2009.ap08746

Abadie, A., & Gardeazabal, J.: The economic costs of conflict: A case study of the Basque Country. Am. Econ. Rev. 93(1), 113-132 (2003) doi:10.1257/000282803321455188

Abadie, A., & L'Hour, J.: A penalized synthetic control estimator for disaggregated data. J. Am. Stat. Assoc. 116(536), 1817-1834 (2021) doi:10.1080/01621459.2021.1971535

Aiken, A. R. A., Starling, J. E., Scott, J. G., & Gomperts, R.: Requests for self-managed medication abortion provided using online telemedicine in 30 US states before and after the Dobbs v Jackson Women's Health Organization Decision. JAMA 328(17), 1768-1770 (2022) doi:10.1001/jama.2022.18865

Antonelli, J., Rubinstein, M., Agniel, D., Smart, R., Stuart, E., Cefalu, M., . . . Griffin, B. A.: Autoregressive models for panel data causal inference with application to state-level opioid policies. Retrieved from http://arxiv.org/abs/2408.09012 2024.

Baker, A., Callaway, B., Cunningham, S., Goodman-Bacon, A., & Sant'Anna, P. H. C. 2025. Difference-in-Differences Designs: A Practitioner's Guide. In: arXiv.

Baker, A. C., Larcker, D. F., & Wang, C. C. Y.: How much should we trust staggered difference-in-differences estimates? Journal of Financial Economics 144(2), 370-395 (2022) doi:10.1016/j.jfineco.2022.01.004

Barsky, B. A., Schnake-Mahl, A., Schmit, C. D., & Burris, S.: Improving the transparency of legal measurement in health policy evaluation—A guide for researchers, reviewers, and editors. JAMA Health Forum 6(3), e250067 (2025) doi:10.1001/jamahealthforum.2025.0067

Ben-Michael, E., Feller, A., & Rothstein, J.: The augmented synthetic control method. J. Am. Stat. Assoc. 116(536), 1789-1803 (2021) doi:10.1080/01621459.2021.1929245

Ben-Michael, E., Feller, A., & Rothstein, J.: Synthetic controls with staggered adoption. J. R. Stat. Soc. B: Statistical Methodology 84(2), 351-381 (2022) doi:10.1111/rssb.12448

Berk, R.A. Causal inference challenges with interrupted time series designs: An evaluation of an assault weapons ban in California. Observational Studies, 8(1), 1-27 (2022). doi: 10.1353/obs.2022.0000

Bernal, J. L., Cummins, S., & Gasparrini, A.: Interrupted time series regression for the evaluation of public health interventions: A tutorial. Int. J. Epidemiol. 46(1), 348-355 (2017) doi:10.1093/ije/dyw098

Bertrand, M., Duflo, E., & Mullainathan, S.: How much should we trust differences-in-differences estimates? The Quarterly Journal of Economics 119(1), 249-275 (2004) doi:10.1162/003355304772839588

Biglan, A., Ary, D., & Wagenaar, A. C.: The value of interrupted time-series experiments for community intervention research. Prev. Sci. 1(1), 31-49 (2000) doi:10.1023/A:1010024016308

Birkland, T. A., Taylor, K., Crow, D. A., & DeLeo, R.: Governing in a polarized era: Federalism and the response of U.S. state and federal governments to the COVID-19 pandemic. Publius: The Journal of Federalism 51(4), 650-672 (2021) doi:10.1093/publius/pjab024

Borusyak, K., Jaravel, X., & Spiess, J.: Revisiting event-study designs: Robust and efficient estimation. Rev. Econ. Stud., rdae007 (2024) doi:10.1093/restud/rdae007

Callaway, B., & Sant'Anna, P. H. C.: Difference-in-Differences with multiple time periods. J. Econom. 225(2), 200-230 (2021) doi:10.1016/j.jeconom.2020.12.001

Chabé-Ferret, S.: Should we combine difference in differences with conditioning on pre-treatment outcomes? TSE Working Paper n. 17-824 (2017).

Ciccarone, D.: The rise of illicit fentanyls, stimulants and the fourth wave of the opioid overdose crisis. Curr. Opin. Psychiatry 34(4), 344-350 (2021) doi:10.1097/YCO.0000000000000717

Dave, D., Friedson, A. I., Matsuzawa, K., & Sabia, J. J.: When do shelter-in-place orders fight Covid-19 best? Policy heterogeneity across states and adoption time. Economic Inquiry 59(1), 29-52 (2021) doi:10.1111/ecin.12944

Daw, J. R., & Hatfield, L. A.: Matching and regression to the mean in difference-in-differences analysis. Health Serv. Res. 53(6), 4138-4156 (2018a) doi:10.1111/1475-6773.12993

Daw, J. R., & Hatfield, L. A.: Matching in difference-in-differences: Between a rock and a hard place. Health Serv. Res. 53(6), 4111 (2018b) doi:10.1111/1475-6773.13017

de Chaisemartin, C., & D'Haultfœuille, X.: Two-Way fixed effects estimators with heterogeneous treatment effects. Am. Econ. Rev. 110(9), 2964-2996 (2020) doi:10.1257/aer.20181169

Degli Esposti, M., Spreckelsen, T., Gasparrini, A., Wiebe, D. J., Bonander, C., Yakubovich, A. R., & Humphreys, D. K.: Can synthetic controls improve causal inference in interrupted time series evaluations of public health interventions? Int. J. Epidemiol. 49(6), 2010-2020 (2020) doi:10.1093/ije/dyaa152

Dimick, J. B., & Ryan, A. M.: Methods for evaluating changes in health care policy: The difference-in-differences approach. JAMA 312(22), 2401-2402 (2014) doi:10.1001/jama.2014.16153

Ewusie, J. E., Soobiah, C., Blondal, E., Beyene, J., Thabane, L., & Hamid, J. S.: Methods, applications and challenges in the analysis of interrupted time series data: A scoping review. J. Multidiscip. Healthc. 13, 411-423 (2020) doi:10.2147/JMDH.S241085

Freyaldenhoven, S., Hansen, C., Pérez, J. P., & Shapiro, J. M.: Visualization, identification, and estimation in the linear panel event-study design. NBER Working Paper Series No. 29170 (2021).

Fry, C. E., & Hatfield, L. A.: Birds of a feather flock together: Comparing controlled pre–post designs. Health Serv. Res. 56(5), 942-952 (2021) doi:10.1111/1475-6773.13697

Gardner, J. 2022. Two-stage differences in differences. In: arXiv.

Goodman-Bacon, A.: Difference-in-differences with variation in treatment timing. J. Econom. 225(2), 254-277 (2021) doi:10.1016/j.jeconom.2021.03.014

Grant, S., Smart, R., & Stein, B. D.: We need a taxonomy of state-level opioid policies. JAMA Health Forum 1(2), e200050 (2020) doi:10.1001/jamahealthforum.2020.0050

Greer, S. L., Dubin, K. A., Falkenbach, M., Jarman, H., & Trump, B. D.: Alignment and authority: Federalism, social policy, and COVID-19 response. Health Policy 127, 12-18 (2023) doi:10.1016/j.healthpol.2022.11.007

Griffin, B. A., Nascimento de Lima, P., Griswold, M., Pane, J., & Grimm, G.: optic: Simulation Tool for Causal Inference Using Longitudinal Data. R package version 1.0.2. Retrieved from https://randcorporation.github.io/optic/ 2025. Accessed 2025 March 27

Griffin, B. A., Schuler, M. S., Pane, J., Patrick, S. W., Smart, R., Stein, B. D., . . . Stuart, E. A.: Methodological considerations for estimating policy effects in the context of co-occurring policies. Health Serv. Outcomes Res. Methodol. 23(2), 149-165 (2023a) doi:10.1007/s10742-022-00284-w

Griffin, B. A., Schuler, M. S., Stone, E. M., Patrick, S. W., Stein, B. D., Nascimento de Lima, P., . . . Stuart, E. A.: Identifying optimal methods for addressing confounding bias when estimating the effects of state-level policies. Epidemiology 34(6), 856 (2023b) doi:10.1097/EDE.0000000000001659

Griffin, B. A., Schuler, M. S., Stuart, E. A., Patrick, S., McNeer, E., Smart, R., . . . Pacula, R. L.: Moving beyond the classic difference-in-differences model: A simulation study comparing statistical methods for estimating effectiveness of state-level policies. BMC Med. Res. Methodol. 21(1), 279 (2021) doi:10.1186/s12874-021-01471-y

Hawkins, N. G., Sanson-Fisher, R. W., Shakeshaft, A., D'Este, C., & Green, L. W.: The multiple baseline design for evaluating population-based research. Am. J. Prev. Med. 33(2), 162-168 (2007) doi:10.1016/j.amepre.2007.03.020

Hernán, M. A., & Robins, J. M.: Using big data to emulate a target trial when a randomized trial is not available. Am. J. Epidemiol. 183(8), 758-764 (2016) doi:10.1093/aje/kwv254

Imai, K., & Kim, I. S.: On the use of two-way fixed effects regression models for causal inference with panel data. Political Analysis 29(3), 405-415 (2021) doi:10.1017/pan.2020.33

Jones, M. R., Novitch, M. B., Sarrafpour, S., Ehrhardt, K. P., Scott, B. B., Orhurhu, V., . . . Simopoulos, T. T.: Government legislation in response to the opioid epidemic. Curr. Pain Headache Rep. 23(6), 40 (2019) doi:10.1007/s11916-019-0781-1

Kontopantelis, E., Doran, T., Springate, D. A., Buchan, I., & Reeves, D.: Regression based quasi-experimental approach when randomisation is not an option: Interrupted time series analysis. BMJ 350(jun09 5), h2750-h2750 (2015) doi:10.1136/bmj.h2750

Kreif, N., Grieve, R., Hangartner, D., Turner, A. J., Nikolova, S., & Sutton, M.: Examination of the synthetic control method for evaluating health policies with multiple treated units. Health Economics 25(12), 1514-1528 (2016) doi:10.1002/hec.3258

Liu, L., Wang, Y., & Xu, Y.: A practical guide to counterfactual estimators for causal inference with time-series cross-sectional data. Am. J. Pol. Sci. 68(1), 160-176 (2024) doi:10.1111/ajps.12723

Lopez Bernal, J., Cummins, S., & Gasparrini, A.: The use of controls in interrupted time series studies of public health interventions. Int. J. Epidemiol. 47(6), 2082-2093 (2018) doi:10.1093/ije/dyy135

Lurie, N., & Sharfstein, J. M.: State-to-state differences in US COVID-19 outcomes: Searching for explanations. Lancet 401(10385), 1314-1315 (2023) doi:10.1016/S0140-6736(23)00726-2

Matthay, E. C., Gottlieb, L. M., Rehkopf, D., Tan, M. L., Vlahov, D., & Glymour, M. M.: What to do when everything happens at once: Analytic approaches to estimate the health effects of co-occurring social policies. Epidemiol. Rev. 43(1), 33-47 (2021) doi:10.1093/epirev/mxab005

Miller, D. L.: An introductory guide to event study models. Journal of Economic Perspectives 37(2), 203-230 (2023) doi:10.1257/jep.37.2.203

Oberlander, J.: The Ten Years' War: Politics, partisanship, and the ACA. Health Aff. 39(3), 471-478 (2020) doi:10.1377/hlthaff.2019.01444

Pacula, R. L., Stein, B. D., Griffin, B. A., Powell, D., & Smart, R.: Improving policy evaluations of the opioid crisis. Health Aff. 44(9), 1042-1049 (2025) doi: 10.1377/hlthaff.2025.00418

Powell, D., Walfon, T., & Griffin, B. A.: Simulating policy effects using lagged values to impute counterfactuals. (Under Review).

Rich, R. F., & White, W. D. 1997. Health Care Policy and the American States: Issues of Federalism. In Health Policy, Federalism and the American States: Routledge.

Roth, J., & Sant'Anna, P. H. C.: Efficient estimation for staggered rollout designs. J. Polit.Econ. Microecon. 1(4), 669-709 (2023) doi:10.1086/726581

Roth, J., Sant'Anna, P. H. C., Bilinski, A., & Poe, J.: What's trending in difference-in-differences? A synthesis of the recent econometrics literature. J. Econom. 235(2), 2218-2244 (2023) doi:10.1016/j.jeconom.2023.03.008

Rudolph, K. E., Gimbrone, C., Matthay, E. C., Díaz, I., Davis, C. S., Keyes, K., & Cerdá, M.: When effects cannot be estimated: Redefining estimands to understand the effects of naloxone access laws. Epidemiology 33(5), 689 (2022) doi:10.1097/EDE.0000000000001502

Ryan, A. M.: Well-balanced or too matchy–matchy? The controversy over matching in difference-in-differences. Health Serv. Res. 53(6), 4106 (2018) doi:10.1111/1475-6773.13015

Ryan, A. M., Burgess Jr, J. F., & Dimick, J. B.: Why we should not be indifferent to specification choices for difference-in-differences. Health Serv. Res. 50(4), 1211-1235 (2015) doi:10.1111/1475-6773.12270

Schell, T. L., Cefalu, M., Griffin, B. A., Smart, R., & Morral, A. R.: Changes in firearm mortality following the implementation of state laws regulating firearm access and use. Proc. Natl. Acad. Sci. 117(26), 14906-14910 (2020) doi:10.1073/pnas.1921965117

Schell, T. L., Griffin, B. A., & Morral, A. R. 2018. Evaluating Methods to Estimate the Effect of State Laws on Firearm Deaths: A Simulation Study (RR-2685-RC). Retrieved from RAND Corporation: https://www.rand.org/pubs/research_reports/RR2685.html

Schuler, M. S., Griffin, B. A., Cerdá, M., McGinty, E. E., & Stuart, E. A.: Methodological challenges and proposed solutions for evaluating opioid policy effectiveness. Health Serv. Outcomes Res. Methodol. 21(1), 21-41 (2021) doi:10.1007/s10742-020-00228-2

Schuler, M. S., Heins, S. E., Smart, R., Griffin, B. A., Powell, D., Stuart, E. A., . . . Stein, B. D.: The state of the science in opioid policy research. Drug Alcohol Depend. 214, 108137 (2020) doi:10.1016/j.drugalcdep.2020.108137

Schuler, M. S., Stuart, E. A., Griffin, B. A., Smart, R., Pacula, R. L., Powell, D., . . . Stein, B. D. 2023. How Much Can You Trust the Results of This Health Policy Evaluation?: A Pragmatic Guide for State Policymakers. Retrieved from https://americanhealth.jhu.edu/sites/default/files/PolicyEvaluationBasics-OPTIC-BAHI-Schuler-March2023.pdf

Seewald, N. J., McGinty, E. E., & Stuart, E. A.: Target trial emulation for evaluating health policy. Ann. Intern. Med. 177(11), 1530-1538 (2024) doi:10.7326/M23-2440

Smart, R., Kase, C. A., Taylor, E. A., Lumsden, S., Smith, S. R., & Stein, B. D.: Strengths and weaknesses of existing data sources to support research to address the opioids crisis. Prev. Med. Rep. 17, 101015 (2020) doi:10.1016/j.pmedr.2019.101015

Snow, J. 1855. *On the Mode of Communication of Cholera*: John Churchill.

Stuart, E. A., Huskamp, H. A., Duckworth, K., Simmons, J., Song, Z., Chernew, M. E., & Barry, C. L.: Using propensity scores in difference-in-differences models to estimate the effects of a policy change. Health Serv. Outcomes Res. Methodol. 14(4), 166-182 (2014) doi:10.1007/s10742-014-0123-z

Sun, L., & Abraham, S.: Estimating dynamic treatment effects in event studies with heterogeneous treatment effects. J. Econom. 225(2), 175-199 (2021) doi:10.1016/j.jeconom.2020.09.006

Wang, G., Hamad, R., & White, J. S.: Advances in difference-in-differences methods for policy evaluation research. Epidemiology 35(5), 628-637 (2024) doi:10.1097/EDE.0000000000001755

Wing, C., Simon, K., & Bello-Gomez, R. A.: Designing difference in difference studies: Best practices for public health policy research. Ann. Rev. Public Health 39(Volume 39, 2018), 453-469 (2018) doi:10.1146/annurev-publhealth-040617-013507

Wooldridge, J. M.: Two-way fixed effects, the two-way Mundlak regression, and difference-in-differences estimators. SSRN Electronic Journal (2021) doi:10.2139/ssrn.3906345

Zeldow, B., & Hatfield, L. A.: Confounding and regression adjustment in difference-in-differences studies. Health Serv. Res. 56(5), 932-941 (2021) doi:10.1111/1475-6773.13666

Zhang, X., & Warner, M. E.: COVID-19 policy differences across US states: Shutdowns, reopening, and mask mandates. Int. J. Environ. Res. Public Health 17(24), 9520 (2020) doi:10.3390/ijerph17249520

**Figure 1.** Methods for policy evaluation

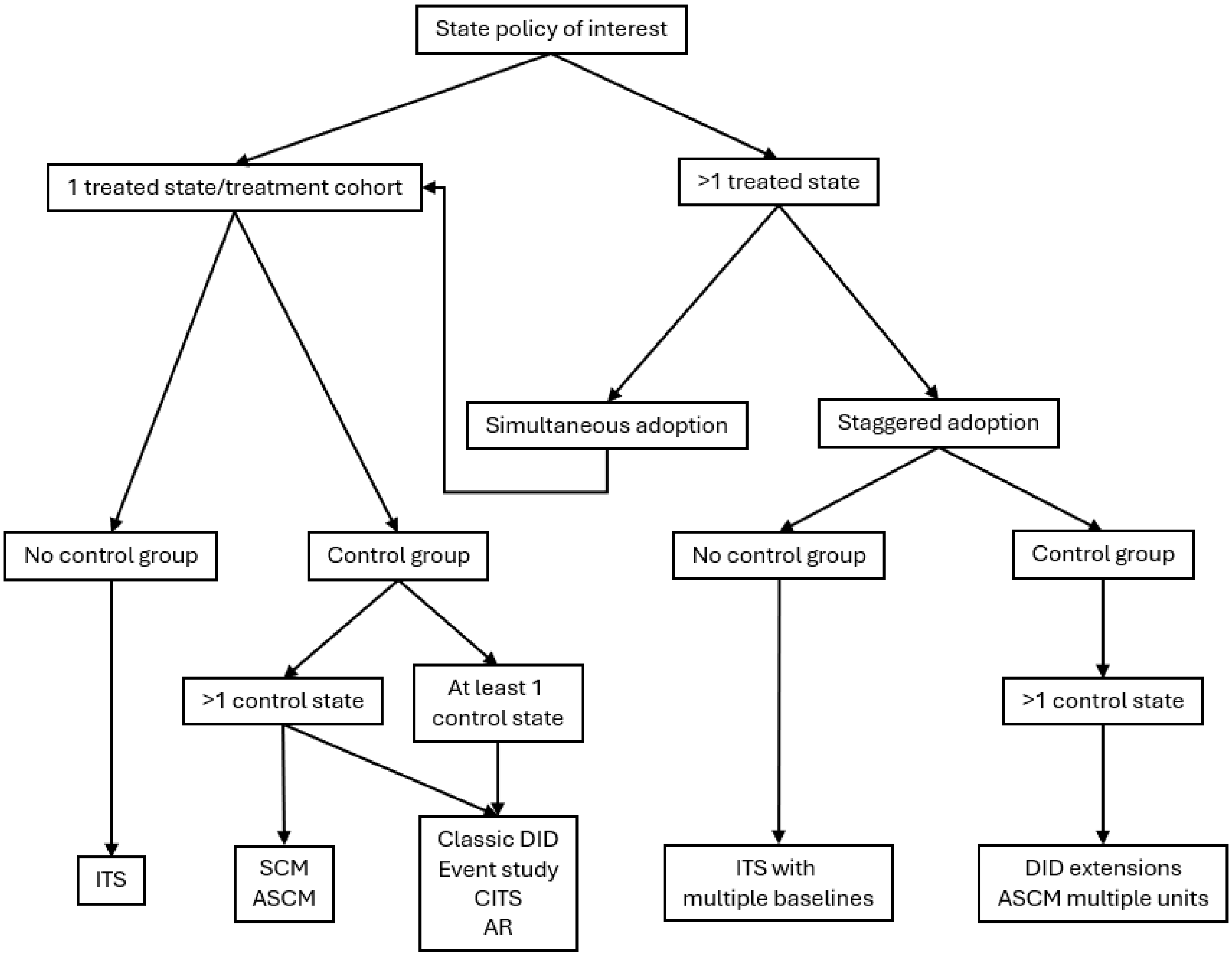

Notes: Figure depicts flow chart for identifying appropriate methods based on number of treatment states, timing of policy adoption, and presence and number of control states. Model assumptions, effect heterogeneity, data considerations, and relative model performance should be assessed and can be compared across models to inform selection of analytic approach.
ITS: interrupted time series; SCM: synthetic control method; ASCM: augmented synthetic control method; DID: difference-in-differences; CITS: comparative interrupted time series; AR: autoregressive

**Table 1.** Common methods for policy evaluation

| Method | Summary | Illustration of modeled policy effects |
|---|---|---|
| ***Methods for a single treated state (or single treatment cohort), no comparison state(s)*** | | |
| Interrupted time series (ITS) | Method that examines repeated outcome measures in a treated unit before and after a policy change. The underlying assumption is that in the absence of a policy, the level and trend of the outcome would have continued along the same trajectory from the pre-period into the post period. | 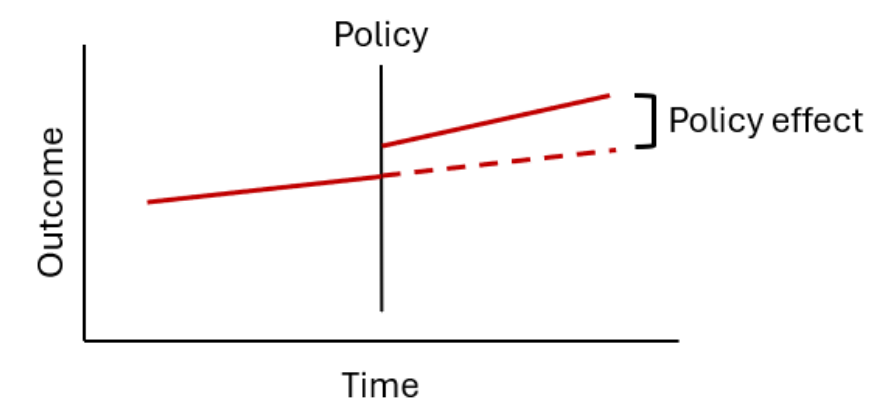  |
| ***Methods for multiple treatment cohorts (staggered adoption), no comparison state(s)*** | | |
| ITS with multiple baselines | Method that extends the ITS approach in settings where multiple units adopt the policy or a single unit adopts and then removes the policy creating multiple baseline periods for pre- and post- comparisons. Similar to ITS, the underlying assumption is that the level and trend of the outcome would have continued along the same trajectory from the pre-period into the post period. | 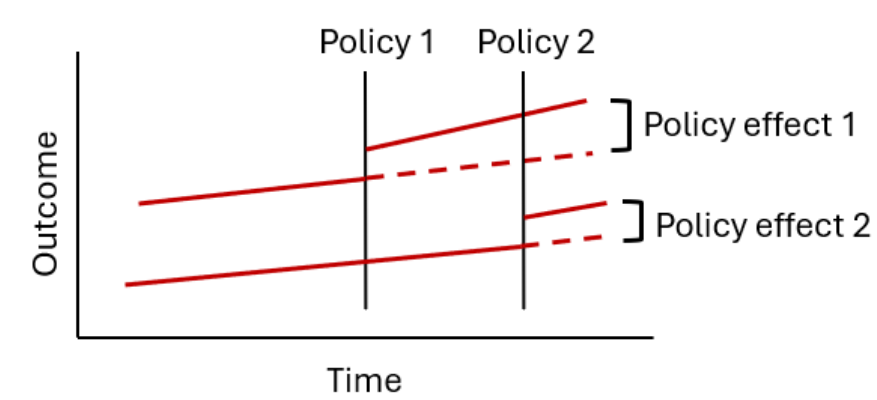  |
| ***Methods for a single treated state, with more than one comparison state*** | | |
| Synthetic control method (SCM) | Method that compares outcomes in a treated unit (e.g., state) to outcomes in a group of control units that has been weighted to best approximate the level and trends of the outcome in the treated unit before policy enactment. The underlying assumption is that the outcomes in the synthetic control are what would have been observed in the treated unit if not for the policy. | 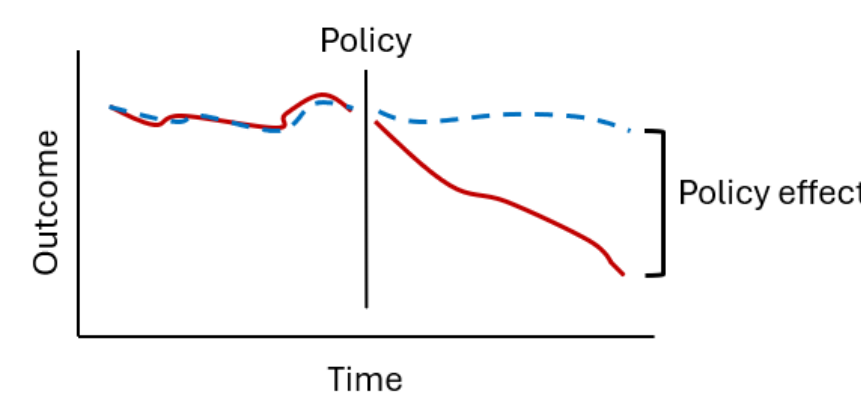  |
| Augmented synthetic control method (ASCM) | The augmented synthetic control method extends the synthetic control method to add an outcome model and allow for negative weighting in the construction of the synthetic control to improve pre-treatment fit. | 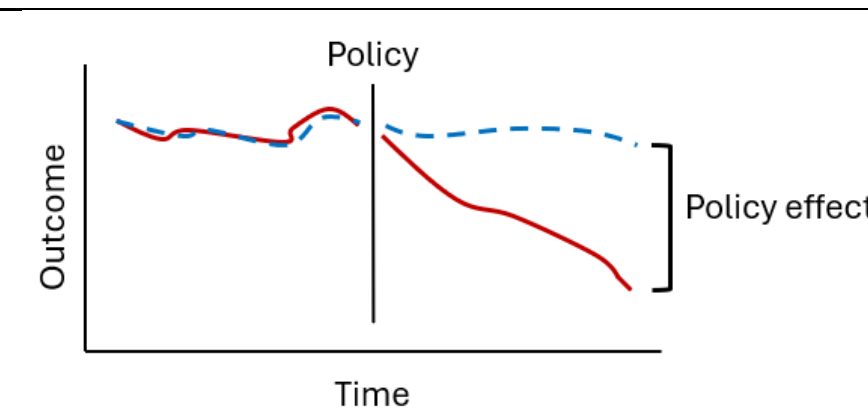  |
| ***Methods for single treated state or treatment cohort (simultaneous adoption), with comparison state(s)*** | | |
| Classic difference-in-differences (DID) | Method that examines pre- and post-treatment outcome measures in treated and control units. The underlying assumption is that in the absence of a policy, the outcome in the treatment units would have changed as much as (or in parallel with) the outcome in the control group. The two-way fixed effects (TWFE) model includes categorical variables for the unit (e.g., state fixed effects) and time (e.g., year fixed effects) to help control for unobserved confounders. | 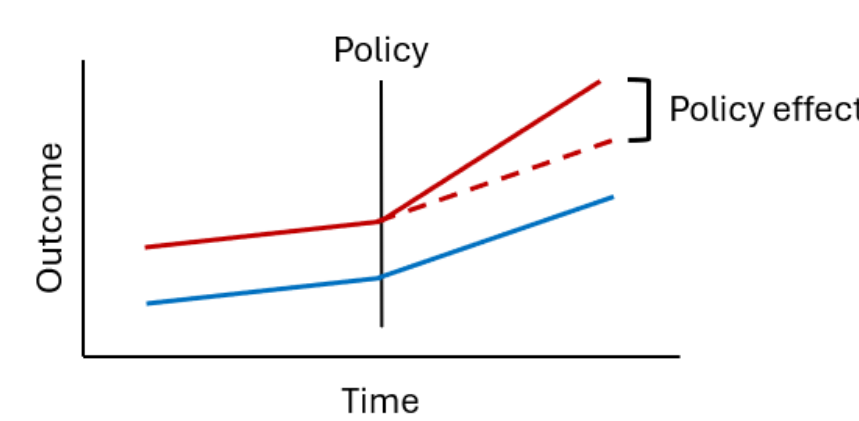  |
| Event study/ Dynamic DID | DID method that estimates a policy effect (comparing treated and control units) at each time point relative to the time of policy enactment by including leading and lagging. | 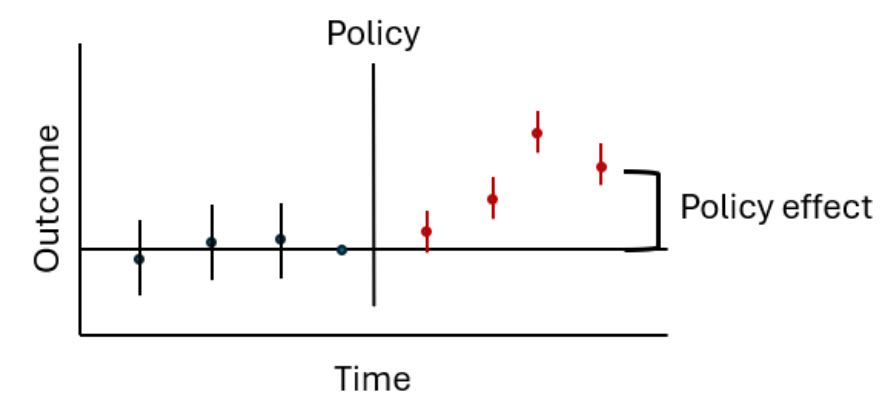  |
| Comparative interrupted time series (CITS) | Method that examines repeated outcome measures in treated and control units before and after a policy change. The underlying assumption is that in the absence of a policy, the level and trend of the outcome in the treated unit would have changed as much as the level and trend in the control group. | 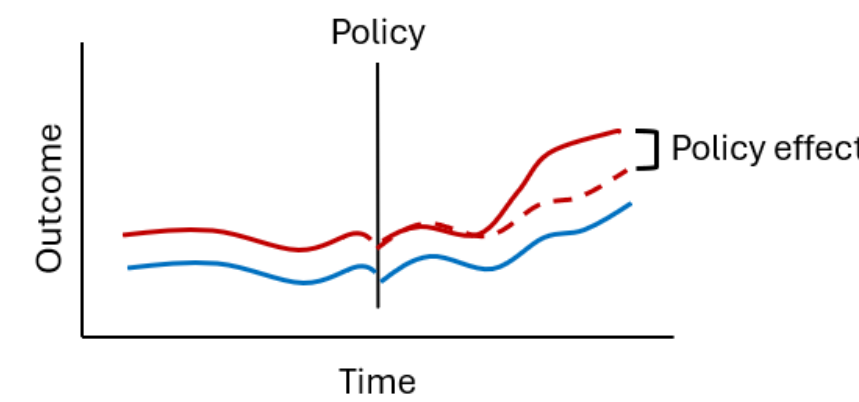  |
| ***Methods for multiple treated states or treatment cohorts (staggered adoption), with comparison state(s)*** | | |
| DID extensions | **Cohort-based DID:** Group of DID methods that account for treatment units adopting the policy and different time points by grouping together units that implemented at the same time and the aggregating effects across all treatment groups.<br><br>**Imputation-based DID:** DID method that accounts for variation in treatment timing by estimating a model using the control/not-yet-treated groups and imputing values for treated groups under control conditions. The difference between the treated and predicted untreated outcomes estimates the treatment effect for each unit-time. These estimates are averaged to form the average treatment effects. | 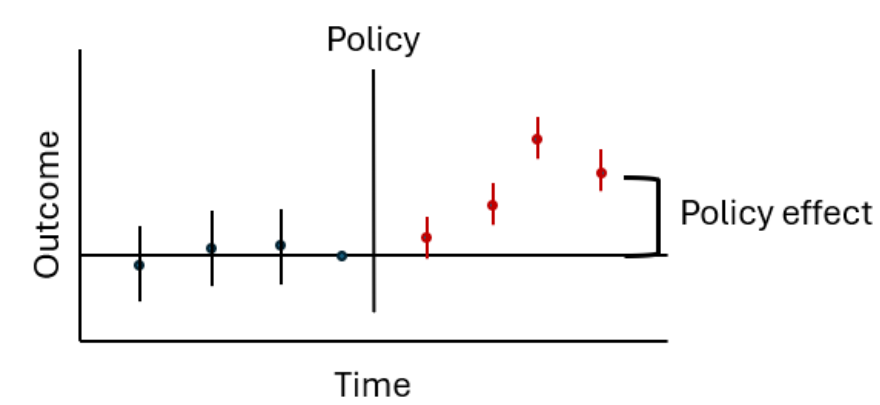  |
| Debiased autoregressive (AR) model | AR regression models include a lagged measure of the outcome to account for correlation between previous and future outcome levels. Debiased AR removes effects of the policies from prior periods as a strategy to obtain unbiased causal effects. The underlying assumption is that the policy is effectively randomized at every time point conditional on the covariates and the prior outcomes absent the policy. | 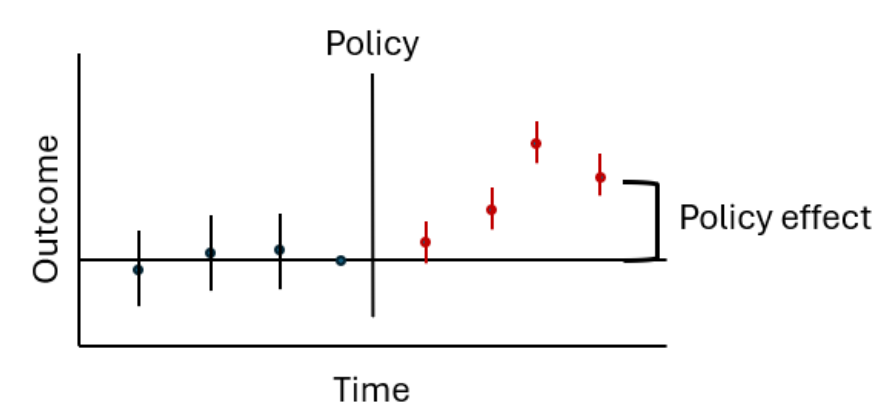  |
| ASCM extension for staggered adoption | Method that extends the augmented synthetic control method for use with multiple treated units by fitting either partially pooled synthetic controls for each treated unit or through combining treated units into treatment time cohorts and fitting a synthetic control for each cohort. The underlying assumption is that the outcomes in the synthetic control are what would have been observed in the treated unit if not for the policy. | 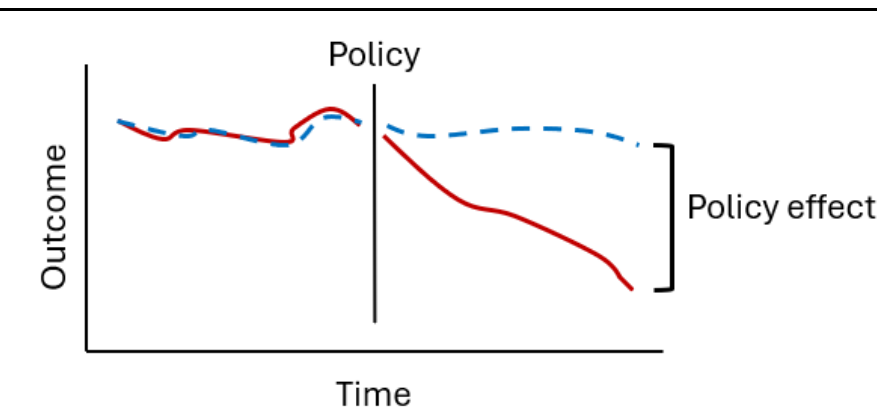  |

**Table 2.** Number of states (including Washington DC) with naloxone standing order laws by year, 1999-2017

| Year | Number of states without policy | Number of states with policy |
|---|---|---|
| 1999 | 51 | 0 |
| 2000 | 51 | 0 |
| 2001 | 51 | 0 |
| 2002 | 51 | 0 |
| 2003 | 51 | 0 |
| 2004 | 51 | 0 |
| 2005 | 51 | 0 |
| 2006 | 51 | 0 |
| 2007 | 51 | 0 |
| 2008 | 51 | 0 |
| 2009 | 51 | 0 |
| 2010 | 50 | 1 |
| 2011 | 50 | 1 |
| 2012 | 50 | 1 |
| 2013 | 49 | 2 |
| 2014 | 41 | 10 |
| 2015 | 33 | 18 |
| 2016 | 17 | 34 |
| 2017 | 7 | 44 |

**Table 3.** Application of study design considerations to case study data

| Study design considerations | Case study scenario |
|---|---|
| Is there a control group? | Yes |
| How many states are available to be used as controls? | 6 never-treated states |
| How many states are in the treated group? | 44 states + Washington DC |
| In the case of multiple treated states, how many "cohorts" are treated? | 6 treatment year cohorts:<br>2010 cohort = 1 state<br>2013 cohort = 1 state<br>2014 cohort = 8 states<br>2015 cohort = 8 states<br>2016 cohort = 16 states<br>2017 cohort = 10 states |
| What is the timing of policy adoption? | Staggered |
| Plausibility of model assumptions? | Potentially different policy provisions across states, potential for spillover |
| Is there potential effect heterogeneity over time? | Yes |
| Is there potential effect heterogeneity by treatment cohort? | Yes |
| Data considerations? | None |
| Additional considerations? | Consider relative model performance |

**Table 4.** Estimated effects of state standing naloxone orders on overall overdose rates by method

| | **Cohort-based DID** | **Imputation-based DID** | **Debiased AR** | **ASCM extension for multiple units** |
|---|---|---|---|---|
| Overall effect estimate<br>(Standard error) | -1.32<br>(6.26) | 2.32<br>(1.35) | 1.47<br>(2.59) | 2.18<br>(2.51) |